# Machine learning meets Singular Optics II: Single-pixel Detection of Structured Light

Purnesh Singh Badavath[†], and Vijay Kumar[*]

Department of Physics, National Institute of Technology Warangal, Telangana, 506004, India

Email: [†]bpurneshsingh@gmail.com, [*]vijay@nitw.ac.in

**Abstract:**

Structured light beams, including Laguerre-Gaussian (LG), Hermite-Gaussian (HG), and perfect vortex (PV) spatial modes, have been at the forefront of modern optics due to their potential in communications, metrology, and sensing. Traditional recognition methods often demand complex alignments and high-resolution imaging. Speckle-learned recognition (SLR) has emerged as a powerful alternative, exploiting the spatio-temporal speckle fields generated by light-diffuser interactions. This paper builds upon the earlier report "Machine Learning Meets Singular Optics" (Proc. SPIE 12938, 2024), which demonstrated structured light recognition using 2D speckle images in both on-axis and off-axis channels captured in the spatial domain. In the present work, the recognition framework is advanced by employing 1D speckle information captured in the spatial and temporal domains. This paper reviews how the 2D spatial information of the structured light is mapped on 1D speckle arrays captured in space and 1D temporal speckle fluctuations recorded in time. The 1D speckle arrays captured in the spatial domain have successfully recognised the parent structured light beams with accuracy exceeding 94%, even by employing 1/nth of the 2D speckle data. More recently, 2D spatial information of structured light beams has been mapped onto temporal speckle sequences recorded by a single-pixel detector in the temporal domain. This study highlights the accuracy exceeding 96% across various structured light families, with resilience to turbulence and modal degeneracy. These advances establish scalable, alignment-free, and low-latency recognition architectures suitable for optical communication, sensing, and quantum technologies.

## 1. Introduction:

Structured light (SL) beams, such as Laguerre-Gaussian ($LG_{p,l}$), Hermite-Gaussian ($HG_{m,n}$), and perfect vortex ($PV_l$) beams are light fields with engineered amplitude, phase, or polarisation distributions. Their ability to provide new degrees of freedom has enabled advances in optical trapping, imaging, metrology, and high-capacity communication systems. For these applications to reach their full potential, reliable recognition of SL beams is essential[1].

Conventional recognition methods, based on diffraction, interference, or modal decomposition, face several challenges. They are highly sensitive to alignment, vulnerable to distortions such as turbulence, and typically require complex optical setups. Machine learning has offered an alternative[2], and in particular, 2D speckle-learned recognition (SLR) has proven powerful technique for structured light detection[3, 4]. By analysing the speckle fields generated when structured beams interact with diffusers, SLR achieves robust classification even under turbulence and noise, even in non-line-of-sight communication channels[5, 6]. Earlier demonstrations established the viability of 2D SLR for a wide range of structured light families along with intensity degenerate beams using astigmatism[7-9]. The OAM beams were also classified using micro and nano-structures from near-field to far-field[10]. However, all of these advances have relied on 2D speckle pattern images. While effective, 2D SLR brings its own limitations: large storage requirements, significant computational cost, and long post-processing times. The reliance on full-frame image capture also adds delay in training and reduces suitability for real-time applications.

To overcome these challenges, SLR has been extended beyond 2D imaging. 1D SLR in space reduces complexity by mapping speckle information to line arrays, preserving recognition accuracy with reduced data[11]. More recently, 0D SLR in the temporal domain has been introduced, where dynamic speckle fluctuations are captured by a single-pixel detector, enabling real-time, low-latency recognition. Together, these approaches pave the way for scalable, efficient, and alignment-free structured light recognition in the spatial and temporal domains[12, 13].

## 2. Spatial 1D Speckle Learned Recognition

When structured light beams interact with a varying or rotating scattering medium such as a diffuser, they generate spatiotemporally varying speckle fields. These speckle patterns, though seemingly random, represent the underlying spatial mode of the incident beam.

In the spatial domain, the speckle grain distribution in a speckle pattern image is isotropically distributed for rotational symmetric beams like $LG_{p,l}$ and $HG_{m,n}$ beams. As the speckle grain distribution is random and isotropic in nature, this allows us to exploit and map the 2D spatial features of the structured light beams on 1D spatial speckle arrays (SSA), as shown in Fig. 1. These 1D SSA retain sufficient statistical features of the original beam to enable accurate recognition. The mapped SSA is shown in Fig. 1.

The $LG_{p,l}$ beams speckle patterns with $p = 0; l = 1 - 8$ have been captured. For each beam, 2000 1D SSA have been mapped from the captured 2D speckle pattern images. The captured 1D SSA will be fed to a custom-designed 1D convolutional neural network (CNN) for training. The trained 1D CNN achieved a peak classification accuracy of 98% when trained on experimental $LG_{p,l}$ 1D SSA of length 1×1920 pixels with 2000 observations per class[11]. The proposed 1D SLR scheme was further applied to $HG_{m,n}$ beams and demonstrated consistent performance with results comparable to those obtained for $LG_{p,l}$ beams.

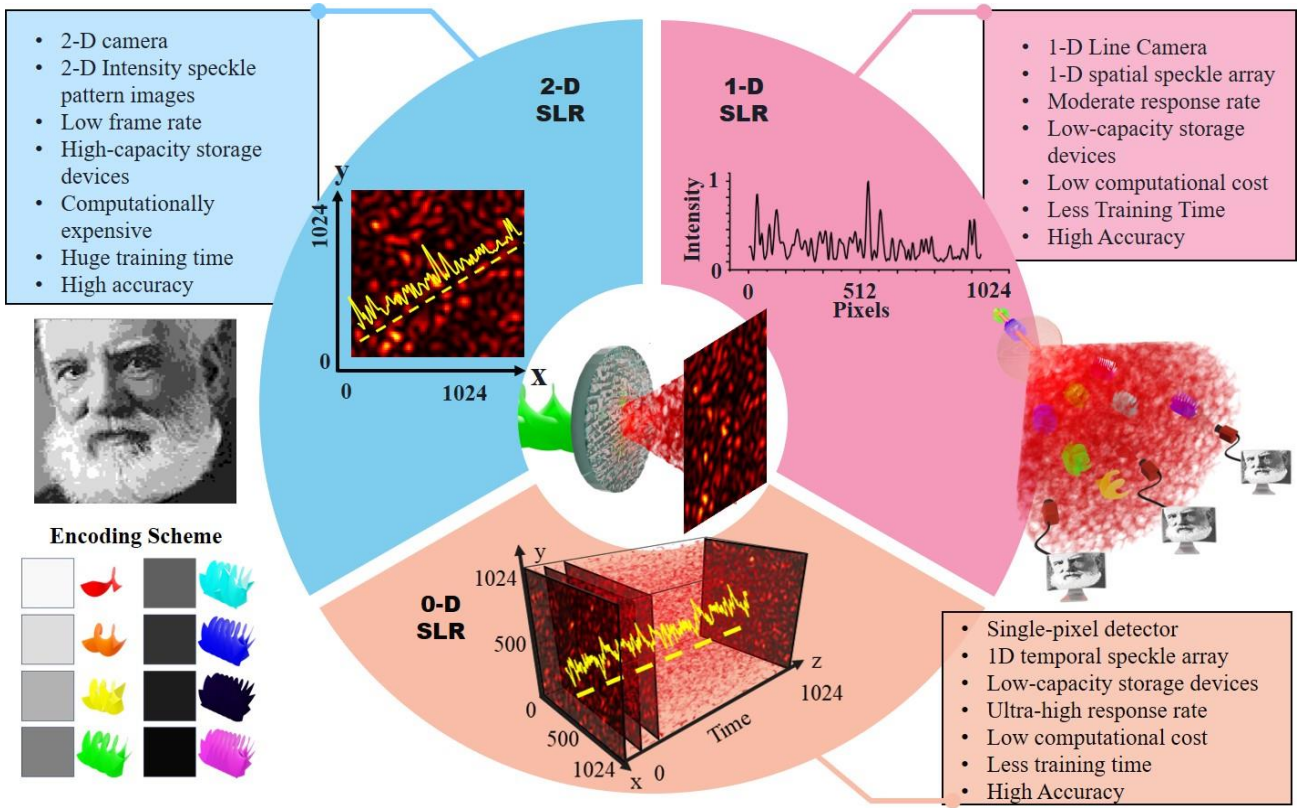

Figure 1. Progression of speckle-learned recognition from 2D image-based detection to 1D spatial arrays and 0D temporal speckle sequences. Each stage reduces storage and computational cost while maintaining high accuracy, enabling scalable and energy-efficient structured light recognition.

## 3. Temporal 0D Speckle learned Recognition

Beyond spatial line mappings, structured light recognition can be further simplified by harnessing the temporal evolution of speckle fields. When the scattering medium is dynamically perturbed, for example by rotating a diffuser, the speckle pattern fluctuates in time. Recording these fluctuations at a point in space, with a single-pixel detector and a photodiode, yields temporal speckle sequences (TSS) that represent the spatial structure of the input beam in the temporal domain. as shown in Fig. 1.

For each beam, 2000 speckle pattern images have been arranged in the sequence they were captured. A $32 \times 32$ single pixel detector grid has been laid out over the simulated and experimentally captured 2D speckle pattern images to map the 1D TSS. For each beam, 1024 1D TSS were mapped. For $LG_{p,l}$ beams with $p = 0$; $l = -8\ to + 8$ (17 beams), a total of 17,408 1D TSS were mapped. For $HG_{m,n}$, $m = n = 1 - 8$ (8 beams), a total of 8,192 1D TSS were mapped. The 1D TSS were fed to a support vector machine (SVM) model for training and testing. The trained SVM model on a $1 \times 1$ pixel has achieved 86.7% and 92.5% accuracies for $LG_{p,l}$ and $HG_{m,n}$ beams[12, 13].

To investigate the influence of speckle grain size relative to detector size, the average speckle grain sizes ($S_S$) were calculated for both the $LG_{p,l}$ and $HG_{m,n}$ mode families. A detector-to-speckle size ratio ($D_S/S_S$) was then established by varying the detector size ($D_S$) from 5.86μm to 164.08μm. The corresponding 1D TSS for each ratio were used to train and test an SVM classifier. The results show that classification accuracy improves as the $D_S/S_S$ ratio increases, owing to the availability of richer speckle information in the mapped sequences. To further quantify this effect, individual accuracies were analysed for different modes. It was observed that beams with $S_S$ larger than the $D_S$ exhibited lower recognition accuracy, whereas cases where the $D_S$ matched or exceeded the $S_S$ resulted in higher accuracy. These findings indicate that for reliable classification, the ratio of detector size to speckle grain size should be $D_S/S_S \geq 1$[13].

## 4. Effect of turbulence on Spatial and Temporal Speckle Fields

In the spatial domain, SLR using both 2D images and 1D line arrays has shown moderate resilience under noisy and turbulent conditions. Experiments demonstrate that even when Kolmogorov turbulence is introduced, classification accuracies remain reasonably high for $LG_{p,l}$ and $HG_{m,n}$ beams. In spatial 1D SLR, where $LG_{p,l}$ beams subjected to the turbulence of the refractive index structure parameter, $C_n^2 = 1.5 \times 10^{-14} m^{\frac{-2}{3}}$, the recognition accuracy was reduced to about 79%, compared to accuracies exceeding 95% in noise-free conditions[11]. The turbulence distorts the beam and introduces asymmetries in beam intensity profiles that weaken the mapping of beam-dependent features, resulting in the degradation of recognition performance.

In contrast, when structured light beams are mapped onto 1D TSS, the dynamic speckle fluctuations encode the spatial information of the structured light beams more effectively. Temporal 0D SLR has demonstrated strong robustness even under severe turbulence up to $C_n^2 = 1 \times 10^{-12} m^{\frac{-2}{3}}$. Reported results show that classification accuracies above 95% can be maintained across $LG_{p,l}$, $HG_{m,n}$, and $PV_l$ families by maintaining the $D_S/S_S$ ratio[13].

A clear advantage of the temporal 0D SLR approach is its independence from asymmetry effects that limit spatial-domain recognition. In spatial SLR (both 2D and 1D), recognition fidelity depends on the isotropy of the sampled speckle field, and turbulence amplifies anisotropy, leading to the accuracy reduction. In contrast, temporal mapping integrates speckle dynamics over time, making it largely immune to turbulence-induced asymmetry. Along with reduced hardware requirements and computational efficiency, this makes temporal 0D SLR a superior solution for practical implementations in free-space optical communication and sensing.

## 4. Summary and Outlook

This work has presented the 1D SLR of structured light beams across spatial and temporal domains. In the spatial domain, 1D SSA mapped from 2D speckle images demonstrated efficient recognition of $LG_{p,l}$ and $HG_{m,n}$ beams. The 1D SLR approach retained the essential mode-dependent features while reducing data size, computational cost, and training time by more than an order of magnitude compared to 2D SLR. Such efficiency makes spatial 1D SLR suitable for applications where storage, energy, and processing resources are constrained.

In the temporal domain, 0D SLR was implemented by mapping TSS with a single-pixel detector. This approach enabled robust recognition of intensity degenerate structured light beams, achieving accuracies above 95% even under severe turbulence conditions. By mapping spatial information into TSS, 0D SLR eliminated the anisotropy and asymmetry sensitivities observed in spatial schemes. With its low-latency operation, independence from imaging sensors, and high resilience to environmental distortions, temporal 0D SLR holds promise for scalable deployment in free-space optical communication and real-time sensing.

Both spatial 1D and temporal 0D SLR approaches are scalable, energy-efficient, and computationally lightweight, enabling faster training and recognition with minimal hardware requirements. They offer sustainable alternatives to traditional 2D camera-based recognition, lowering the overall power consumption and system complexity while preserving high fidelity in beam classification.

Looking ahead, these methods open multiple future directions. In telecommunications, they can enable high-capacity, high-speed, turbulence-resilient free-space optical links. In sensing, single-pixel speckle-based architectures may be extended to biomedical diagnostics, material inspection, and metrology. Furthermore, their inherent energy efficiency and low computation cost align with the growing need for sustainable photonic technologies, suitable for satellite links, remote communication in energy-limited environments, and integration into compact on-chip photonic devices. By bridging singular optics with machine learning across spatial and temporal dimensions, this work underscores a new paradigm of structured light recognition that is robust, fast, and sustainable, paving the way for next-generation optical systems in both research and industry.

**Acknowledgement:** We acknowledge Prof. Nirmal K. V. for allowing us to use the lab facility.